\documentstyle[epsf,12pt]{article}

\newcommand{\mb}[1]{ { \mbox{\boldmath{$#1$}}}  }

\begin{document}

\begin{center} 
{\Large Three Band Model for Superconductivity  in Sr$_2$RuO$_4$} \\~ \\
G. Litak$^1$, J.F. Annett$^2$, B.L. Gy\"{o}rffy$^2$, K.I. Wysoki\'nski$^3$  \\~
\\ $^1$Department of Mechanics, Technical University of Lublin, \\
Nadbystrzycka 36,
20-618 Lublin, Poland \\
$^2$H.H. Wills Physics Laboratory, University of Bristol, Tyndall
Ave, Bristol, BS8 1TL, UK\\
$^3$Institute of Physics, Maria Curie-Sk\l{}odowska University, \\
Radziszewskiego 10, PL-20-031 Lublin, Poland 

\end{center}


\begin{abstract} 
We analyse the pairing symmetry of the order parameter in  superconducting
Sr$_2$RuO$_4$.
Basing on a realistic three dimensional 3 band energy spectrum 
 we have introduced effective attractive electron-electron  
 interactions and  found a solution
which has  line nodes in the gap in $\alpha$ and $\beta$ bands
but not in $\gamma$ band.
This state  breaks 
time reversal symmetry and leads to the proper temperature dependence 
of the specific heat, in-plane penetration depth and thermal conductivity.  
\end{abstract}

\section{Introduction} 
 Sr$_2$RuO$_4$ is an oxide with a lattice structure similar to
the well known high temperature superconductor La$_{2-x}$Ba$_x$CuO$_4$
\cite{Mae94,Mae01} with ruthenium replacing copper. When pure it becomes a 
superconductor at T$_c$=1.5K and 
 is one of the best 
candidates  for spin triplet  superconductivity \cite{Mae01,Bas96,Agt97}.
 
In spite of many convincing arguments for $p$--wave pairing  the proper
symmetry of three band superconducting order parameter is still unknown.

The muon spin rotation experiments \cite{Luk98} indicate the
existence time reversal symmetry breaking, consistent with
the order parameter
$\hat{\Delta}({\bf k})=
 i\hat{\sigma}_y\hat{{\mathbf \sigma}}\cdot{\bf d}({\bf k})$,
with ${\bf d}({\bf k})=(0,0,d^z({\bf k}))$ and 
 $d^z({\bf k})=\Delta(T)(k_x+ik_y)$. At the same time many thermodynamic
properties show power low temperature dependences at low T, clearly pointing at
the existence of nodes in the order parameter. 
The experiments on specific heat \cite{Nis00}, penetration depth
\cite{Bon00} and thermal conductivity \cite{Iza01}
all
show a quadratic dependence on temperature. 
This is surprising since a complete symmetry analysis of $p$--wave pairing in tetragonal 
crystals shows no states which
both break time reversal
and possess line nodes \cite{Ann90,Sig99}.

In the literature there exist a number of proposals for the relevant 
symmetry states
\cite{Bas96,Agt97,Spa01,Miy99,Maz97,Tew99,Has00,Zhi01}. 
Most of them concentrate on two dimensional band models with
p-wave symmetry states. Here we propose to start with the real three dimensional
electron energy spectrum with all three bands (called $\alpha$, 
$\beta$ and $\gamma$) taken into account and allow for both in plane and out
plane p-wave symmetries
of the order parameter. 

 In the next section  we present our Hamiltonian and the approach, which is
 the full solution of the Bogolubov - de Gennes equations at each point of the
 three dimensional Brillouin zone. In principle our theory does not contain free
 parameters. By fitting the calculated Fermi surface to the 
 experimentally determined one  we fixed the single particle parameters of the
 Hamiltonian. The remaining three interaction parameters have been fitted by
 requiring that all three bands have a single superconducting transition
 temperature 
 (i.e. the monotonic low T behaviour of the specific heat) and a correct
 value of the transition temperature. Using the parameters we have solved the
 Bogolubov-de Gennes equations.   The obtained gap functions have been used to
 calculate penetration depth and thermal conductivity.

\section{The Symmetry of the Order Parameter}

We start with the 
description of  the electronic structure of  Sr$_2$RuO$_4$ in the vicinity of
the Fermi energy $\epsilon_F$ in the normal state. 
We consider three orbitals of ruthenium A, B, C of character $d_{xz}$, $d_{yz}$ and 
$d_{xy}$, respectively. These generate 3 Fermi surface sheets $\alpha$, $\beta$
and $\gamma$. The $\alpha$ and $\beta$ sheets arise from hybridised  A and B
orbitals which we model with hopping parameters ($t_1$, $t_2$, $t_{AB}$), 
which describe AA (BB) hopping in x(y) direction, AA (BB) hopping in y(x)
direction and AB hopping, respectively. 
 The $\gamma$ sheet arises form the C orbitals, which we model by hopping
 parameters $t$ and $t^{\prime}$ for nearest and next nearest neighbour sites.
We have fitted these parameters to the de Haas - van Alphen experimental
cross-sections of the Fermi surface \cite{Mac96}. To allow for the three
dimensional Fermi surface \cite{Ber00} we add the hopping parameter $t_{\perp}$ which we
assume to take on the same value for each band.

\begin{figure}[htb]
\vspace{-2cm}
\hspace*{0.5cm}
\epsfxsize=4.0cm
\epsffile{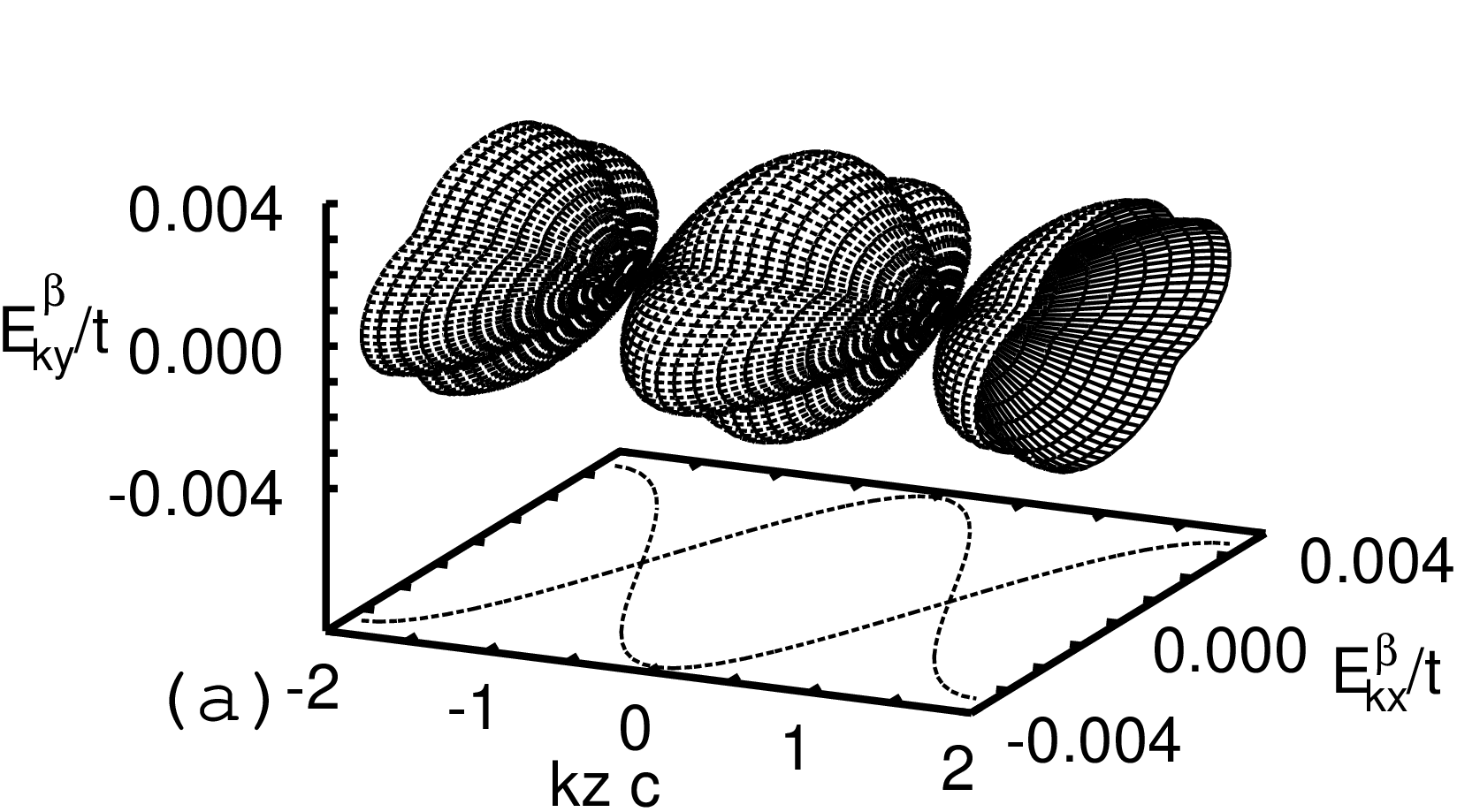}
\hspace*{2.0cm}
\epsfxsize=4.0cm
\epsffile{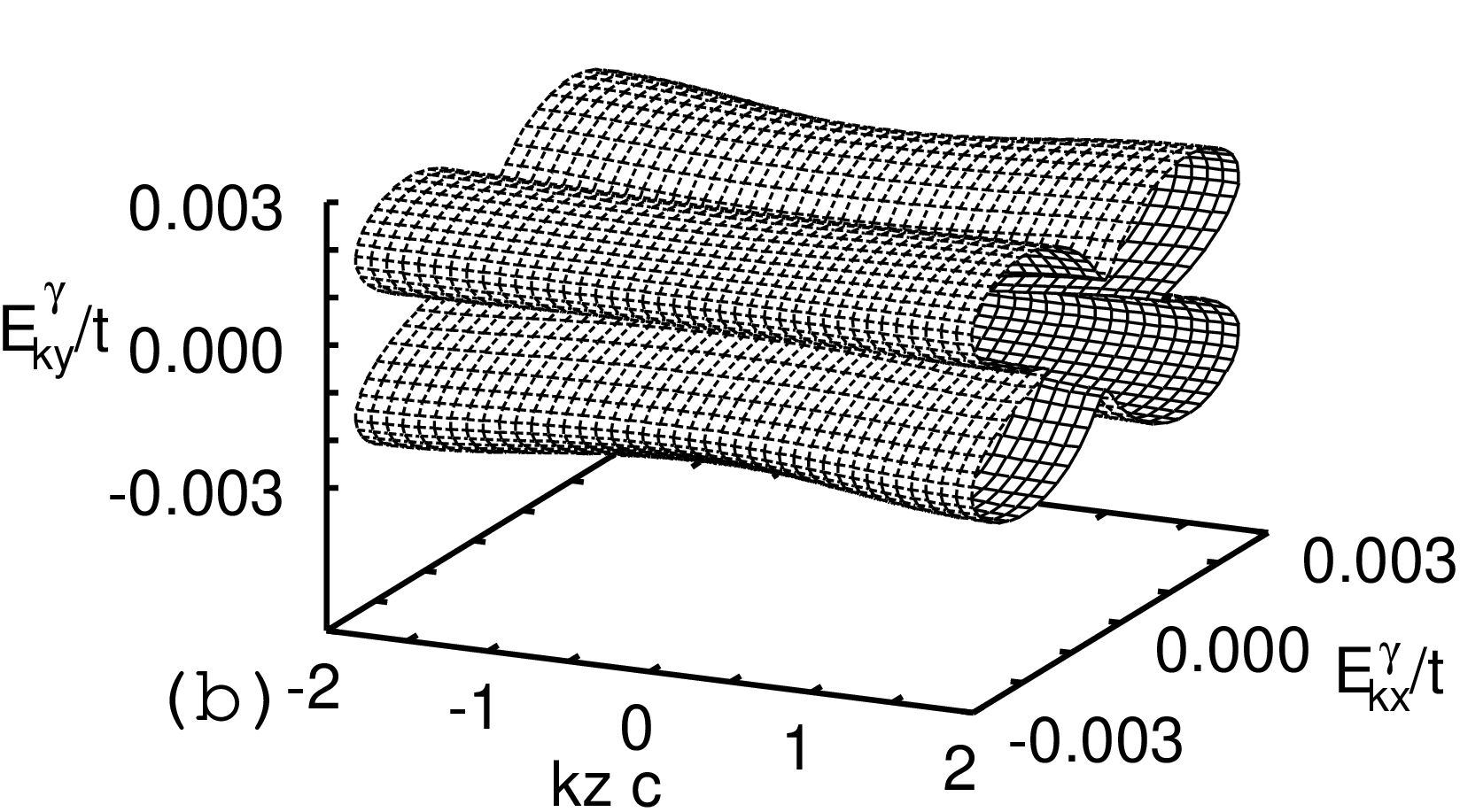}
\vspace*{-1.5cm}

\vspace{2cm}
\caption[Fig. 1]{Eigenvalues  $E^{\beta, \gamma}(\mb
k)$ at the 
$\beta$
(a) and $\gamma$ (c)  Fermi surfaces
plotted as a function of $k_z$ and xy plane azimuthal angle.}
\end{figure}

Due to the  symmetry of the orbitals and the body  centered tetragonal 
symmetry of
the crystal  we assumed that the dominant pairing interactions were 
in plane $U_{\gamma}$ for the $\gamma$ band, and inter-plane $U_{\alpha,\beta}$ 
for the $\alpha$ and $\beta$ bands.
Each of these was 
tuned by fitting to the experimental superconducting critical
temperature $T_C=1.5$ K, so we have no free parameters.

Our strategy is to solve the Bogolubov-de Gennes equations
to find the pairing potentials and energy spectrum.
We allowed for the following general structure of the
pairing potential 

\begin{eqnarray}
d^z_{\nu,\nu\prime}(\bf k) &=&\Delta_x^{\nu,\nu\prime} \sin{k_x}
+\Delta_y^{\nu,\nu\prime} \sin{k_y}  \nonumber \\
&+&\left[\Delta_{x\prime}^{\nu,\nu\prime} \sin{(k_x/2)} \cos{(k_y/2)}  
+\Delta_{y\prime}^{\nu,\nu\prime} \sin{(k_y/2)} \cos{(k_x/2)} \right] 
\cos{(k_z c/2)}
\nonumber \\
&+&\Delta_{z\prime}^{\nu,\nu\prime} \sin{(k_zc/2)} \cos{(k_x/2)} \cos{(k_y/2)}, 
 \end{eqnarray}
in units where the lattice constant, $a=1$, and where
indices $\nu$, $\nu\prime$ take the values $\alpha$, $\beta$, $\gamma$.

\begin{figure}[htb]
\leavevmode
\vspace{-1.5cm}

\hspace*{0.5cm}
\epsfxsize=4.6cm
\epsffile{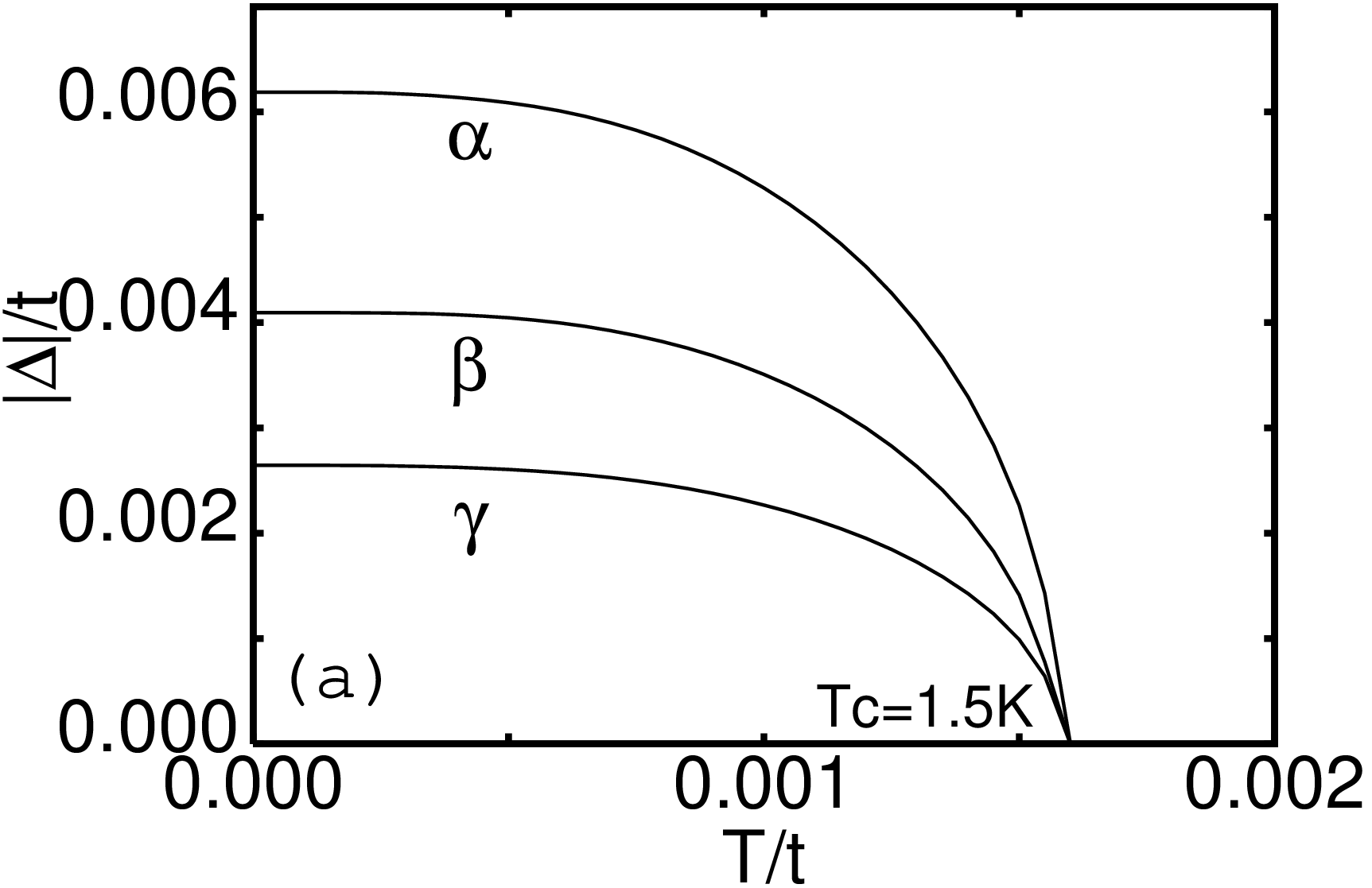}

\vspace*{-4.5cm}
\hspace*{9.5cm}
\epsfxsize=4.8cm
\epsffile{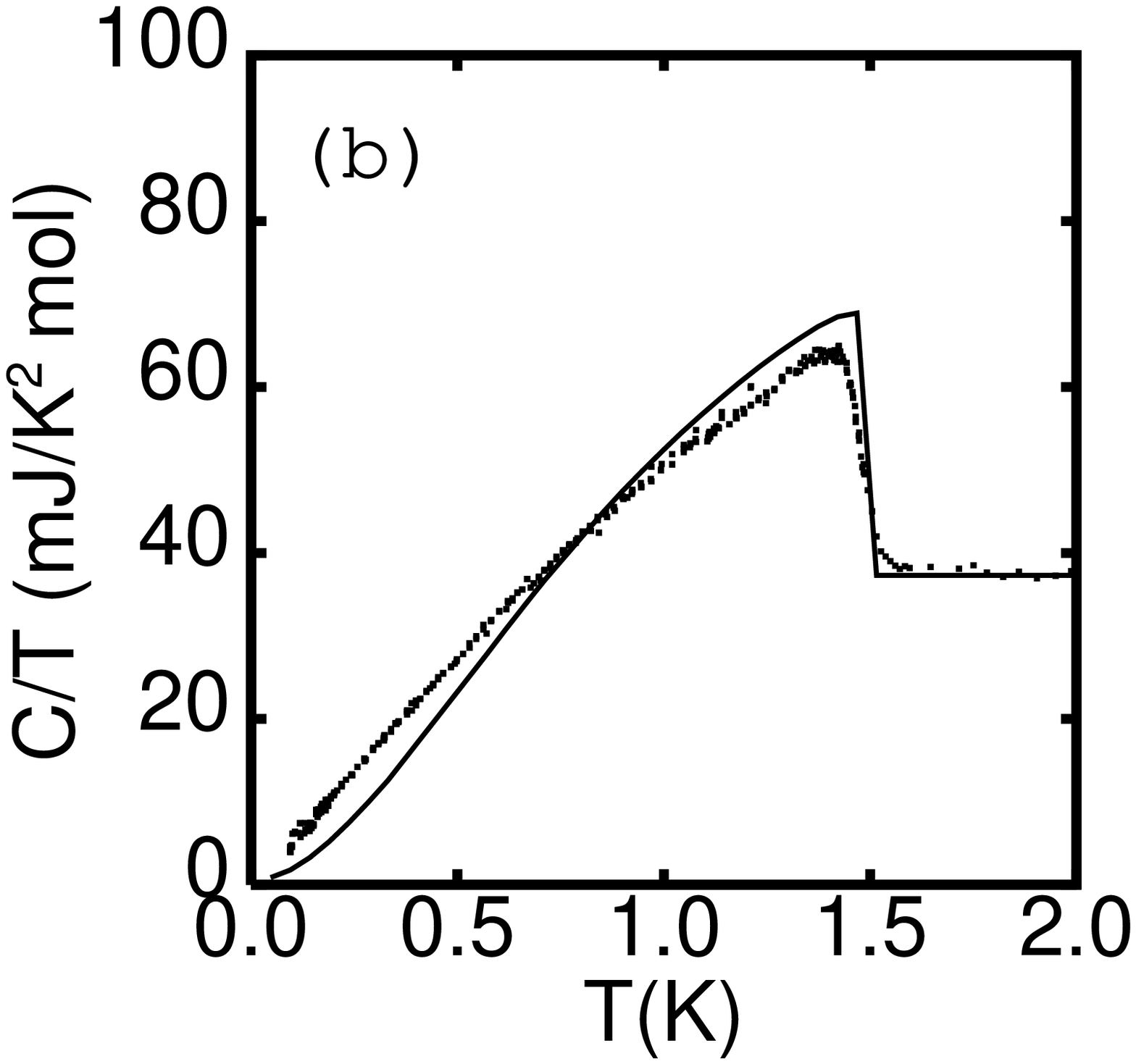}
\vspace{-1.0cm}
\caption[Fig. 2]{ (a)
Pairing potentials $|\Delta|=|\Delta_{x}^{\alpha, \beta, \gamma}|=
|\Delta_{y}^{\alpha, \beta, \gamma}|$ for the three bands $\alpha, \beta,
\gamma$ as a function of temperature. (b) The specific heat $C/T$ versus
temperature (the solid line corresponds to the calculated results,
while the points are the experimental data \cite{Nis00}). }
\end{figure}

\begin{figure}[htb]
\leavevmode
\vspace{1cm}

\hspace*{2.0cm}
\epsfxsize=4.9cm
\
\epsffile{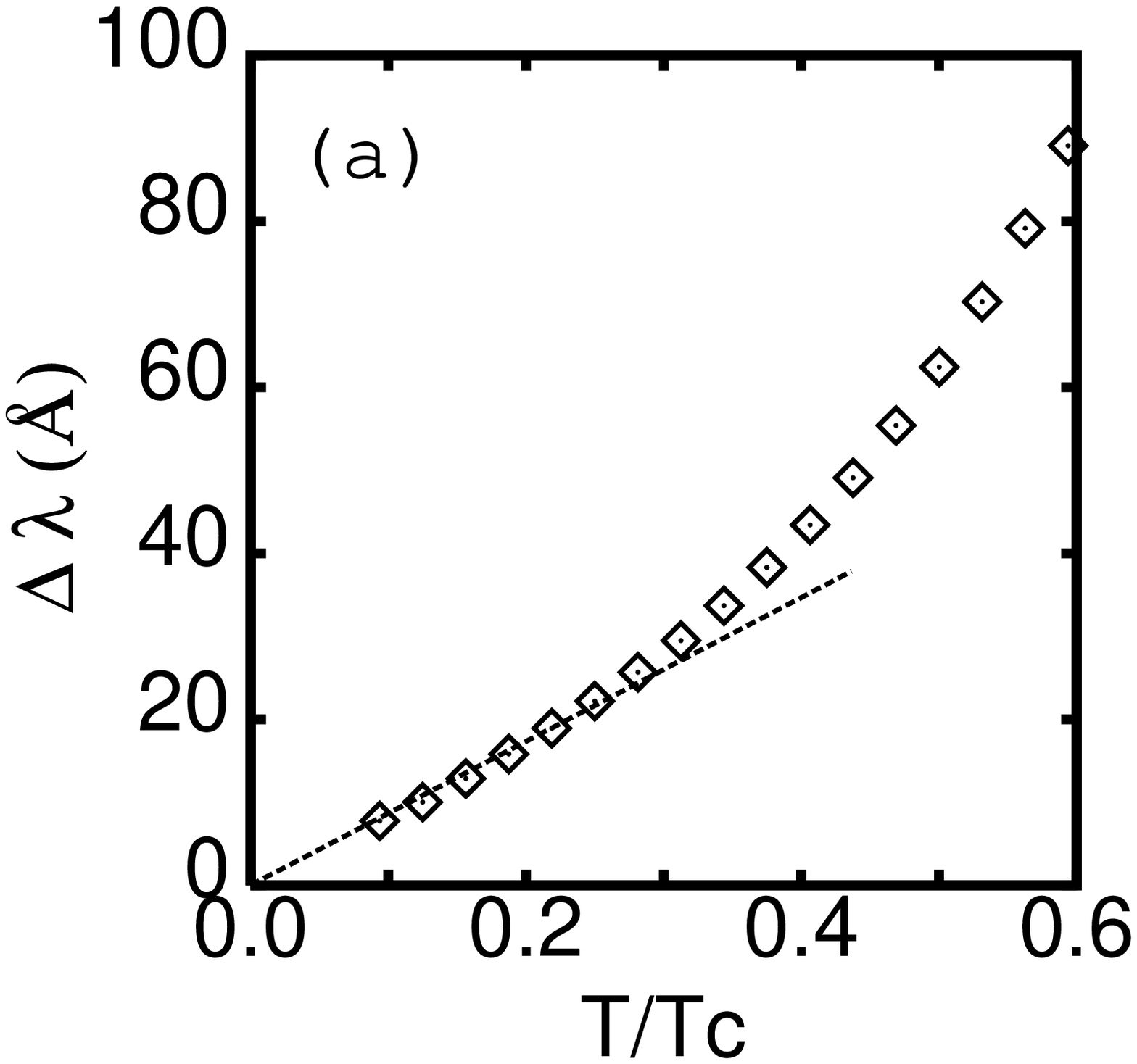}
\hspace*{1.2cm}
\epsfxsize=4.9cm
\epsffile{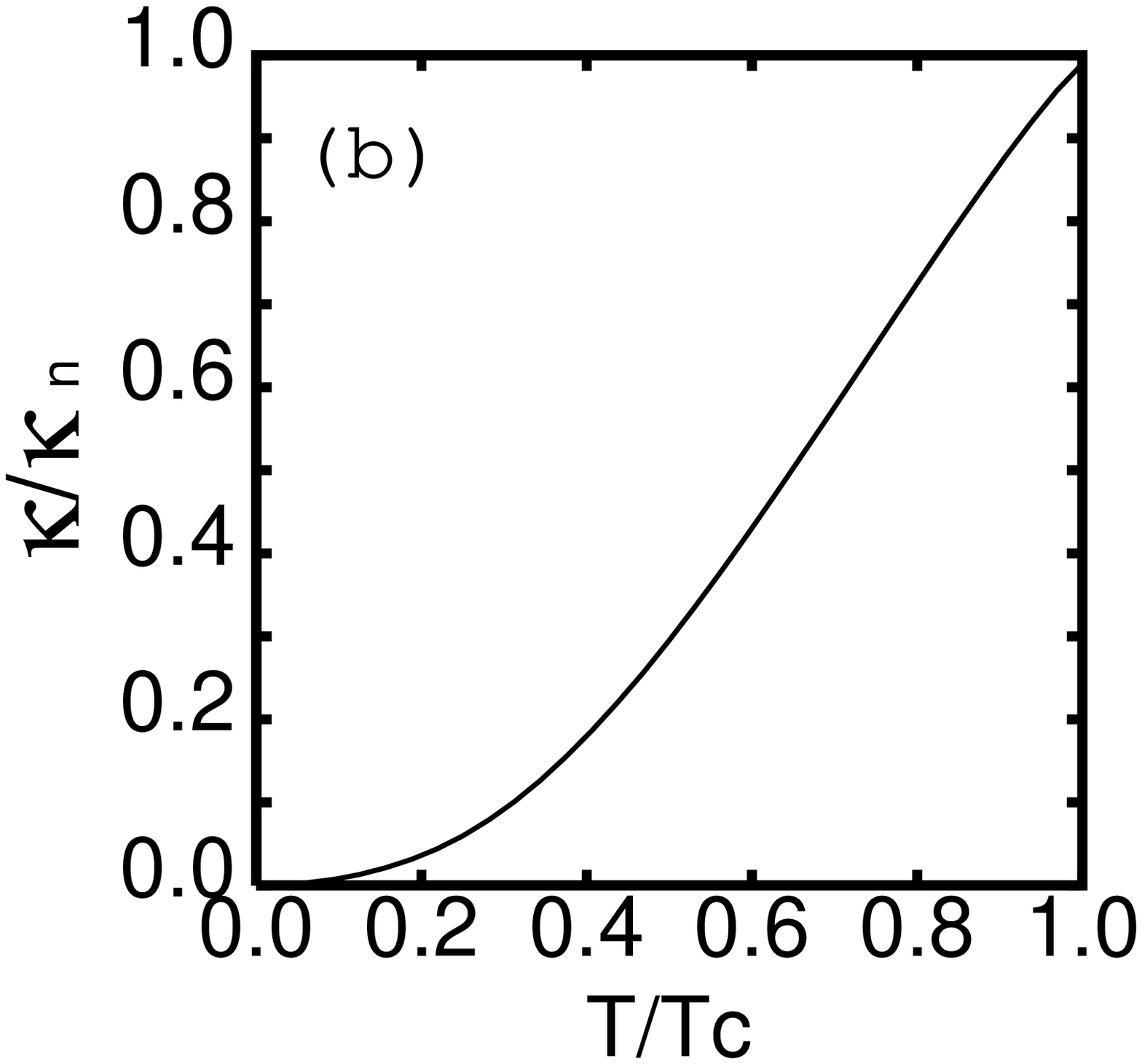}
\vspace{-1cm}
\caption[Fig. 3]{The in plane penetration depth $\lambda$ for low
temperature  (a) (the diamonds
correspond to  the results of calculations while  the solid line is a
linear fit to them).
The thermal conductivity $\kappa$ as
function of
temperature (b).}
\end{figure}

 For  the given set of parameters the only nonzero
 pairing potentials have the following symmetry
 $\Delta^{\alpha,\alpha}_{x\prime} = i\Delta^{\alpha,\alpha}_{y\prime}$, 
$\Delta^{\beta,\beta}_{x\prime}=i \Delta^{\beta,\beta}_{y\prime} $
$\Delta^{\gamma,\gamma}_x= i \Delta^{\gamma,\gamma}_y $.
Each of these solutions therefore breaks the time-reversal symmetry.
Moreover, due to the  factor $\cos(k_z c/2)$ (Eq. 1) we have planes of zeros on
the $\alpha$ and $\beta$ sheets
for $k_z= \pm \pi/c$.  

To illustrate the structure of the gap on the Fermi surface 
we have plotted in Fig. 1 the eigenvalues for the $\beta$ and
$\gamma$ sheets.
Figure (1a) illustrates clearly the zero eigenvalues in the $\beta$ band 
on the planes $k_z= \pm \pi/c$.  The structure of the gap on the $\alpha$ sheet
is very similar. Contrary to that the gap on the
 $\gamma$ band (Fig. 1b) is always finite but has strong modulation in the (x,y) plane.
 
The temperature dependence of the gap parameters is shown in Fig. 2a and the
specific heat in Fig. 2b. The overall agreement with the experimental data 
is good. In particular the height of the step at $T_c$ and the linear dependence
of $C/T$ on temperature agree with experiment \cite{Nis00}.

For the above  state we have also calculated the penetration depth $\lambda$ 
(Fig. 3a) and
the thermal conductivity $\kappa$ (Fig. 3b).
The penetration depth shows a close to linear dependence on temperature 
 at low $T$ and agrees approximately 
with the experimental data \cite{Bon00}. Note that impurity scattering or
other effects may change the low temperature power low.
The thermal conductivity is quadratic as expected for a gap with line nodes.

\section{Remarks and Conclusions}

Using a realistic three band model of Sr$_2$RuO$_4$  we have found a $p$--wave
pairing state which has  nodes and breaks time reversal symmetry.
The state has $^3E_u[e]$ (axial) pairing symmetry in the notation of  \cite{Ann90}.
The line node on the $\alpha$, $\beta$ sheets is a consequence of the specific form 
assumed for the pairing interaction.

The state we have found is of similar nature as the one proposed on different
grounds in \cite{Has00,Zhi01}. Our model contains only attractive
interactions unlike one proposed in \cite{Has00}. The nice feature of our
model is that we have no adjustable parameters. The state we obtained leads 
to correct temperature dependence of not only specific heat but also the
penetration depth and thermal conductivity.

\section*{Acknowledgements}
This work has been partially supported by KBN grant No. 2P03B 106 18 and the 
Royal Society Joint Project. We are grateful to Prof. Y. Maeno for
providing us with 
the experimental data used in Fig. 2b.

\end{document}